\documentclass[12pt]{article}
\usepackage{amsmath}
\usepackage{amssymb}
\usepackage{amsfonts}

\title{Revisiting generalized Hulth\'en potentials}
\author{C. Quesne\thanks{e-mail: cquesne@ulb.ac.be}\\ 
{\small\sl Physique Nucl\'eaire Th\'eorique et Physique Math\'ematique,  Universit\'e Libre de Bruxelles,} \\ 
{\small\sl Campus de la Plaine CP229, Boulevard~du Triomphe, B-1050 Brussels, Belgium}}
\date{ }
\begin{document}
\baselineskip=22pt plus 1pt minus 1pt
\maketitle
\begin{abstract}
A relation between the deformed Hulth\'en potential and the Eckart one is used to write the bound-state wavefunctions of the former in terms of Jacobi polynomials and to calculate their normalization coefficients. The shape invariance property of the Eckart potential in standard first-order supersymmetric quantum mechanics allows to easily rederive the set of extended deformed Hulth\'en potentials, recently obtained by using the Darboux-Crum transformation, and to show that their spectra and normalized wavefunctions follow without any further calculation. The present approach considerably simplifies the previous derivation. Furthermore, by taking advantage of other known rational extensions of the Eckart potential obtained in first-order supersymmetric quantum mechanics, novel extensions of the deformed Hulth\'en potential are constructed, together with their bound-state spectra and wavefunctions. These new extensions belong to three different types, the first two being isospectral to some previously obtained extensions and the third one with an extra bound state below their spectrum.
\end{abstract}


%
%
\newpage
\section{Introduction}

The Hulth\'en potential, which was initially introduced to describe the deuteron \cite{hulthen42, hulthen57, flugge}, has found a lot of applications in nuclear, particle, atomic, and condensed matter physics (see, {\sl e.g.}, \cite{durand, vandijk, gruninger, szalewicz, lindhard}). This potential is a short-range one, which behaves like a Coulomb potential for small values of the radial variable $r$ and decreases exponentially for large values of $r$. The corresponding Schr\"odinger equation can be solved in a closed form for $s$ waves, while, for $l\ne 0$, a number of methods have been employed to find approximate solutions (see, {\sl e.g.}, \cite{lai, patil, popor, roy, hall92}). A deformed version of the Hulth\'en potential has also been proposed \cite{egrifes} for some physical applications.\par
%
%
In a recent work \cite{hall18}, Hall, Saad, and Sen calculated the exact normalized eigenfunctions of the Schr\"odinger equation for such a deformed Hulth\'en potential. They also constructed some extensions of the latter by means of the Darboux-Crum transformation and determined the exact solutions of these extensions.\par
%
%
The purpose of the present paper is twofold.\par
%
%
{}First, we will rederive the results of Ref.~\cite{hall18} by using a simpler alternative approach, based on the Hulth\'en potential property of being a limiting form of the Eckart potential \cite{cooper}. Since the latter is known to be a shape invariant potential in first-order supersymmetric quantum mechanics (SUSYQM) \cite{cooper, genden}, which means that its partner in standard unbroken SUSYQM is similar in shape and differs only in the parameters that appear in it, a whole hierarchy of exactly solvable Eckart potentials with known eigenvalues and eigenfunctions can be directly constructed and leads in the limit to the set of extended deformed Hulth\'en potentials derived in \cite{hall18}.\par
%
%
Second, we will take advantage of a similar type of approach to build some new extensions of the deformed Hulth\'en potential. This time, instead of using the ground-state wavefunction of the starting potential as factorization function in SUSYQM, we will employ polynomial-type, nodeless solutions of the Eckart potential Schr\"odinger equation with an energy below the ground state. As previously shown \cite{cq}, the resulting extensions belong to three different types, the first two being strictly isospectral to an Eckart potential with different parameters and the third one having an extra bound state below the spectrum of the latter. The translation of these results for the deformed Hulth\'en potential will then directly provide us with three new types of extensions.\par
%
%
The paper is organized as follows. In sec.~2, we review the description of the deformed Hulth\'en potential from some known results for the Eckart one. In sec.~3, we derive the extended deformed Hulth\'en potentials obtained by deleting the ground state of the starting potential. In sec.~4, more general extended deformed Hulth\'en potentials are presented. Finally, some concluding remarks are given in sec.~5.\par
%
%
\section{The deformed Hulth\'en potential as a limiting form of the Eckart one}

The Schr\"odinger equation for the deformed Hulth\'en potential can be written as \cite{hall18}
\begin{equation}
  \left[- \frac{1}{2} \frac{d^2}{dx^2} + V_q(x) - E_n\right] \psi_n(x) = 0,  \label{eq:SE}
\end{equation}
where
\begin{equation}
  V_q(x) = - \frac{\mu e^{-\delta x}}{1 - q e^{-\delta x}}, \qquad \frac{\log q}{\delta} < x < +\infty,
  \label{eq:hulthen-pot}
\end{equation}
and $\mu$, $\delta$, $q$ are three positive constants. The constant $\mu$ can be related to the atomic number when the potential is used for atomic phenomena, $\delta$ is a screening parameter that determines the potential range, and $q$ is a deforming parameter, which reduces to one for the standard Hulth\'en potential.\par
%
%
On making the change of variable
\begin{equation}
  r = \delta x - \log q, \qquad \text{where} \qquad 0 < r < +\infty,  \label{eq:r-x}
\end{equation}
implying $e^{-r} = q e^{-\delta x}$, the differential equation (\ref{eq:SE}) can be transformed into
\begin{equation}
  \left[- \frac{d^2}{dr^2} + \bar{V}(r) - \bar{E}_n\right] \bar{\psi}_n(r) = 0,
\end{equation}
where
\begin{equation}
  \bar{V}(r) = - \bar{\mu} \frac{e^{-r/2}}{e^{r/2}-e^{-r/2}}, \qquad \bar{\mu} = \frac{2\mu}{q\delta^2},
  \label{eq:def-V-bar}
\end{equation}
and 
\begin{equation}
  \bar{E}_n = \frac{2E_n}{\delta^2}, \qquad \bar{\psi}_n(r) = \frac{1}{\sqrt{\delta}} \psi_n(x).  \label{eq:def-E-bar}
\end{equation}
\par
%
%
As observed in \cite{cooper}, the potential $\bar{V}(r)$ can be rewritten as
\begin{equation}
  \bar{V}(r) = - \frac{\bar{\mu}}{2} \left(\coth\frac{r}{2} - 1\right)
\end{equation}
and can therefore be considered, up to an additive constant $\bar{\mu}/2$, as a limiting form of the Eckart potential
\begin{equation}
  U_{A,B}(r;\alpha) = A(A-\alpha) \operatorname{csch}^2 \alpha r - 2B \coth \alpha r, \qquad A \ge \alpha, 
  \qquad B>A^2,  \label{eq:Eckart-pot}
\end{equation}
namely
\begin{equation}
  \bar{V}(r) = U_{\frac{1}{2},\frac{\bar{\mu}}{4}}\left(r; \frac{1}{2}\right) + \frac{\bar{\mu}}{2}.
  \label{eq:V-bar}
\end{equation}
Note that, when dealing with the Eckart potential, the condition $A>\alpha$ is often assumed ensuring that the potential is repulsive for $r\to 0$. In the limit $A\to\alpha$, the potential behaviour at the origin is changed, but the results for bound states remain valid. From the Schr\"odinger equation for the Eckart potential
\begin{equation}
  \left[- \frac{d^2}{dr^2} + U_{A,B}(r;\alpha) - {\cal E}_n\right] \phi_n(r) = 0,  \label{eq:Eckart-SE}
\end{equation}
we shall therefore get
\begin{equation}
  \bar{E}_n = {\cal E}_n + \frac{\bar{\mu}}{2}, \qquad \bar{\psi}_n(r) = \phi_n(r),
\end{equation}
for $A=\alpha=1/2$ and $B=\bar{\mu}/4$.\par
%
%
As well known \cite{cooper}, the potential (\ref{eq:Eckart-pot}) has a finite number of bound states with energies
\begin{equation}
  {\cal E}_n = - (A+n\alpha)^2 - \frac{B^2}{(A+n\alpha)^2}, \qquad 0 \le n < \frac{\sqrt{B}-A}{\alpha},
  \label{eq:Eckart-E}
\end{equation}
and wavefunctions
\begin{equation}
  \phi_n(r) = {\cal N}_n (z-1)^{\alpha_n/2} (z+1)^{\beta_n/2} P_n^{(\alpha_n,\beta_n)}(z),
\end{equation}
where
\begin{equation}
  z = \coth \alpha r, \quad \alpha_n = \frac{1}{\alpha}\left(-A-n\alpha + \frac{B}{A+n\alpha}\right), \quad
  \beta_n = \frac{1}{\alpha}\left(-A-n\alpha - \frac{B}{A+n\alpha}\right),  \label{eq:alpha-beta}
\end{equation}
$P_n^{(\alpha_n,\beta_n)}(z)$ is a Jacobi polynomial, and ${\cal N}_n$ a normalization constant.\par
%
%
The calculation of ${\cal N}_n$ can be easily performed by using the standard expansion of Jacobi polynomials \cite{abramowitz}
\begin{equation}
  P_n^{(\alpha_n,\beta_n)}(z) = \frac{1}{2^n} \sum_{m=0}^n \binom{\alpha_n+n}{m} \binom{\beta_n+n}
  {n-m} (z-1)^{n-m} (z+1)^m,  \label{eq:jacobi}
\end{equation}
where here $\alpha_n+n > 0$ and $\beta_n+n < 0$. The second binomial coefficient on the right-hand side of (\ref{eq:jacobi}) can be transformed into
\begin{align}
  \binom{\beta_n+n}{n-m} &= \frac{\Gamma(\beta_n+n+1)}{(n-m)! \Gamma(\beta_n+m+1)} = (-1)^{n-m}
  \frac{\Gamma(-\beta_n-m)}{(n-m)! \Gamma(-\beta_n-n)} \nonumber \\
  & = (-1)^{n-m} \binom{-\beta_n-m-1}{n-m},
\end{align}
where now $-\beta_n-m-1>0$. On introducing such an expansion twice into
\begin{equation}
  {\cal N}_n^{-2} = \int_0^{\infty} dr\, (\coth\alpha r - 1)^{\alpha_n} (\coth\alpha r + 1)^{\beta_n}
  \left[P_n^{(\alpha_n,\beta_n)}(\coth\alpha r)\right]^2,
\end{equation}
we are left with the integral
\begin{align}
  & \int_0^{\infty} dr\, (\coth\alpha r - 1)^{\alpha_n+2n-m-m'} (\coth\alpha r + 1)^{\beta_n+m+m'} 
       \nonumber \\
  & = \int_0^{\infty} dr\, (\sinh\alpha r)^{-\alpha_n-\beta_n-2n} e^{-\alpha(\alpha_n-\beta_n+2n-2m-2m')r},
\end{align}
which can be performed by using eq.~(3.541.1) of Ref.~\cite{gradshteyn}. The normalization coefficient is therefore given by the following double finite sum
\begin{align}
  {\cal N}_n &= 2^{-\frac{1}{2}(\alpha_n+\beta_n)+\frac{1}{2}} \sqrt{\alpha} \biggl\{\sum_{m,m'=0}^n
       (-1)^{m+m'} \binom{\alpha_n+n}{m} \binom{-\beta_n-m-1}{n-m} \binom{\alpha_n+n}{m'} \nonumber\\
  &{} \times \binom{-\beta_n-m'-1}{n-m'} \frac{\Gamma(\alpha_n+2n-m-m') \Gamma(-\alpha_n-
       \beta_n-2n+1)}{\Gamma(-\beta_n-m-m'+1)}\biggr\}^{-1/2}.  \label{eq:Eckart-norm}
\end{align}
\par
%
%
{}From these results, we infer that the potential $\bar{V}(r)$ of eq.~(\ref{eq:V-bar}) has the bound-state energies
\begin{equation}
  \bar{E}_n = - \frac{1}{4} \left(\frac{\bar{\mu}-(n+1)^2}{n+1}\right)^2, \qquad 0 \le n < \sqrt{\bar{\mu}}
  - 1,
\end{equation}
with corresponding wavefunctions
\begin{align}
  \bar{\psi}_n(r) &= \bar{N}_n \left(\coth\frac{r}{2}-1\right)^{\frac{1}{2}\left(\frac{\bar{\mu}}{n+1}-n-1
       \right)} \left(\coth\frac{r}{2}+1\right)^{-\frac{1}{2}\left(\frac{\bar{\mu}}{n+1}+n+1\right)}
       \nonumber \\
  &{}\quad \times P_n^{\left(\frac{\bar{\mu}}{n+1}-n-1, - \frac{\bar{\mu}}{n+1}-n-1\right)}\left(\coth\frac{r}  
       {2}\right), \\
  \bar{N}_n &= 2^{n+\frac{1}{2}} \biggl\{\sum_{m,m'=0}^n (-1)^{m+m'} \binom{\frac{\bar{\mu}}{n+1}-1}{m}
       \binom{\frac{\bar{\mu}}{n+1}+n-m}{n-m} \binom{\frac{\bar{\mu}}{n+1}-1}{m'} \nonumber \\
  &{}\quad \times \binom{\frac{\bar{\mu}}{n+1}+n-m'}{n-m'} \frac{\Gamma\left(\frac{\bar{\mu}}{n+1}+n-m-
       m'-1\right)}{\Gamma\left(\frac{\bar{\mu}}{n+1}+n-m-m'+2\right)} \biggr\}^{-1/2}.
\end{align} 
\par
%
%
{}For the starting equation (\ref{eq:SE}), from (\ref{eq:hulthen-pot}), (\ref{eq:r-x}), (\ref{eq:def-V-bar}), and (\ref{eq:def-E-bar}), by some straightforward calculations we get the results
\begin{equation}
  E_n = - \frac{1}{2} \left(\frac{\mu}{q\delta(n+1)} - \frac{\delta}{2}(n+1)\right)^2, \qquad 0 \le n <
  \sqrt{\frac{2\mu}{q\delta^2}} - 1,  \label{eq:E}
\end{equation}
and
\begin{align}
  \psi_n(x) &= N_n e^{-\left(\frac{\mu}{q\delta(n+1)} - \frac{1}{2}(n+1)\delta\right)x} \left(1 - qe^{-\delta x}
       \right)^{n+1} \nonumber \\
  &{}\quad \times P_n^{\left(\frac{2\mu}{q\delta^2(n+1)}-n-1, -\frac{2\mu}{q\delta^2(n+1)}-n-1\right)}
       \left(\frac{1+qe^{-\delta x}}{1-qe^{-\delta x}}\right)
\end{align}
with
\begin{equation}
  N_n = \sqrt{\delta} \bar{N}_n 2^{-n-1} q^{\frac{1}{2}\left(\frac{2\mu}{q\delta^2(n+1)}-n-1\right)},
\end{equation}
thence
\begin{align}
  N_n &= \sqrt{\frac{\delta}{2}} q^{\frac{1}{2}\left(\frac{2\mu}{q\delta^2(n+1)}-n-1\right)} 
       \Biggl\{\sum_{m,m'=0}^n (-1)^{m+m'} \binom{\frac{2\mu}{q\delta^2(n+1)}-1}{m} \nonumber \\
  &{}\quad \times \binom{\frac{2\mu}{q\delta^2(n+1)}+n-m}{n-m} \binom{\frac{2\mu}{q\delta^2(n+1)}-1}{m'}  
       \binom{\frac{2\mu}{q\delta^2(n+1)}+n-m'}{n-m'} \nonumber \\
  &{} \quad \times \frac{\Gamma\left(\frac{2\mu}{q\delta^2(n+1)}+n-m-m'-1\right)}
       {\Gamma\left(\frac{2\mu}{q\delta^2(n+1)}+n-m-m'+2\right)} \Biggr\}^{-1/2}.
\end{align}
As previously observed in \cite{hall18}, we note from eq.~(\ref{eq:E}) that for the potential $V_q(x)$ to have bound states, the deforming parameter must be restricted to values $q < 2\mu/\delta^2$.\par
%
%
\section{Extended deformed Hulth\'en potentials obtained by standard unbroken SUSYQM}

Let us start from the Hamiltonian of the Eckart potential (\ref{eq:Eckart-pot}),
\begin{equation}
  {\cal H}_0 = - \frac{d^2}{dr^2} + U^{(0)}(r), \qquad U^{(0)}(r) = U_{A,B}(r;\alpha),
\end{equation}
and define the superpotential
\begin{equation}
  W_{A,B}(r;\alpha) = - \frac{d}{dr}(\log \phi_0(r)) = - A \coth\alpha r + \frac{B}{A},
\end{equation}
in terms of the ground-state wavefunction
\begin{align}
  \phi_0(r) &\propto (\coth\alpha r-1)^{\frac{1}{2\alpha}\left(-A+\frac{B}{A}\right)}
       (\coth\alpha r+1)^{-\frac{1}{2\alpha}\left(A+\frac{B}{A}\right)} \nonumber \\
   &\propto (\sinh\alpha r)^{\frac{A}{\alpha}} e^{- \frac{B}{A}r}.     
\end{align}
Then $U^{(0)}(r)$ can be rewritten as
\begin{equation}
  U^{(0)}(r) = W_{A,B}^2 - \frac{dW_{A,B}}{dr} + {\cal E}_0, \qquad \text{where} \qquad {\cal E}_0 = - A^2
  - \frac{B^2}{A^2},
\end{equation}
and the supersymmetric partner of $U^{(0)}(r)$ is
\begin{equation}
  U^{(1)}(r) = W_{A,B}^2 + \frac{dW_{A,B}}{dr} + {\cal E}_0.
\end{equation}
As well known \cite{cooper}, $U^{(1)}(r)$ is another Eckart potential, namely
\begin{equation}
  U^{(1)}(r) = U_{A+\alpha,B}(r;\alpha).
\end{equation}
\par
%
%
Such a shape invariance property allows one to construct a SUSY hierarchy of Hamiltonians
\begin{equation}
  {\cal H}_i = - \frac{d^2}{dr^2} + U^{(i)}(r), \qquad U^{(i)}(r) = U_{A+i\alpha,B}(r;\alpha), \qquad 
  i=1, 2, 3, \ldots,
\end{equation}
where
\begin{equation}
  U^{(i)}(r) = W_{A+i\alpha,B}^2 - \frac{dW_{A+i\alpha,B}}{dr} + {\cal E}_i = W_{A+(i-1)\alpha,B}^2 +
  \frac{dW_{A+(i-1)\alpha,B}}{dr} + {\cal E}_{i-1}, 
\end{equation}
and ${\cal E}_i$, ${\cal E}_{i-1}$ are given in (\ref{eq:Eckart-E}). Note that since $U^{(0)}(r)$ has a finite number of bound states, the resulting SUSY hierarchy is also a finite one. The bound-state spectrum of ${\cal H}_i$ is given by
\begin{equation}
  {\cal E}^{(i)}_n = {\cal E}^{(0)}_{n+i} = - [A+(n+i)\alpha]^2 - \frac{B^2}{[A+(n+i)\alpha]^2}, \qquad
  0 \le n < \frac{\sqrt{B}-(A+i\alpha)}{\alpha},
\end{equation}
and the corresponding wavefunctions are
\begin{equation}
  \phi^{(i)}_n(r) = {\cal N}^{(i)}_n (z-1)^{\alpha_{n+i}/2} (z+1)^{\beta_{n+i}/2} P_n^{(\alpha_{n+i},
  \beta_{n+i})}(z),
\end{equation}
where $\alpha_{n+i}$ and $\beta_{n+i}$ are defined as in eq.~(\ref{eq:alpha-beta}), but with $n$ replaced by $n+i$. A similar substitution of $\alpha_{n+i}$ and $\beta_{n+i}$ for $\alpha_n$ and $\beta_n$ provides ${\cal N}^{(i)}_n$ from ${\cal N}^{(0)}_n = {\cal N}_n$, given in (\ref{eq:Eckart-norm}).\par
%
%
On assuming now $A=\alpha=1/2$ and $B=\bar{\mu}/4$, we get from $\bar{V}(r)$, given in eqs.~(\ref{eq:def-V-bar}) and (\ref{eq:V-bar}), a set of potentials
\begin{equation}
  \bar{V}^{(i)}(r) = U^{(i)}_{\frac{1}{2},\frac{\bar{\mu}}{4}}\left(r;\frac{1}{2}\right) + \frac{\bar{\mu}}{2} = 
  \frac{1}{4}i(i+1) \operatorname{csch}^2 \frac{r}{2} - \frac{1}{2}\bar{\mu} \coth\frac{r}{2} + \frac{1}{2}
  \bar{\mu}, \qquad i=0,1,2,\ldots,  \label{eq:def-V-bar-i}
\end{equation}
which can be rewritten as
\begin{equation}
  \bar{V}^{(i)}(r) = [i(i+1)-\bar{\mu}] \frac{e^{-r/2}}{e^{r/2} - e^{-r/2}} + \frac{i(i+1) e^{-r}}
  {(e^{r/2}-e^{-r/2})^2}.  \label{eq:V-bar-i}
\end{equation}
Their bound-state energies and wavefunctions are given by
\begin{equation}
  \bar{E}^{(i)}_n = - \frac{1}{4}\left(\frac{\bar{\mu}-(n+i+1)^2}{n+i+1}\right)^2, \qquad 0 \le n <
  \sqrt{\bar{\mu}}-i-1,  \label{eq:E-bar-i}
\end{equation}
and
\begin{align}
  \bar{\psi}^{(i)}_n(r) &= \bar{N}^{(i)}_n \left(\coth\frac{r}{2}-1\right)^{\frac{1}{2}\left(\frac{\bar{\mu}}
       {n+i+1}-n-i-1\right)} \left(\coth\frac{r}{2}+1\right)^{-\frac{1}{2}\left(\frac{\bar{\mu}}
       {n+i+1}+n+i+1\right)} \nonumber \\
  &{}\quad \times P_n^{\left(\frac{\bar{\mu}}{n+i+1}-n-i-1, -\frac{\bar{\mu}}{n+i+1}-n-i-1\right)}
       \left(\coth\frac{r}{2}\right),  \label{eq:psi-bar-i} \\
  \bar{N}^{(i)}_n &= \bar{N}_{n+i},  \label{eq:N-bar-i}
\end{align}
respectively.\par
%
%
The changes of variable (\ref{eq:r-x}) and of parameter (\ref{eq:def-V-bar}) then directly provide us with the Schr\"odinger equation
\begin{equation}
  \left[- \frac{1}{2} \frac{d^2}{dx^2} + V^{(i)}_q(x) - E^{(i)}_n\right] \psi^{(i)}_n(x) = 0, \qquad 
  i=0,1,2,\ldots,  \label{eq:gen-SE}
\end{equation}
for the generalized deformed Hulth\'en potentials
\begin{equation}
  V^{(i)}_q(x) = - \frac{\mu e^{-\delta x}}{1 - qe^{-\delta x}} + \frac{1}{2}i(i+1) \frac{q\delta^2
  e^{-\delta x}}{\left(1 - qe^{-\delta x}\right)^2},  \label{eq:gen-hulthen}
\end{equation}
derived in Ref.~\cite{hall18}. Their corresponding bound-state energies and wavefunctions follow from eqs.~(\ref{eq:E-bar-i}), (\ref{eq:psi-bar-i}), (\ref{eq:N-bar-i}), and are given by
\begin{equation}
  E^{(i)}_n =  - \frac{1}{2}\left(\frac{\mu}{q\delta(n+i+1)} - \frac{\delta}{2}(n+i+1)\right)^2, \qquad
  0 \le n < \sqrt{\frac{2\mu}{q\delta^2}}-i-1, 
\end{equation}
and
\begin{align}
  \psi^{(i)}_n(x) &= N^{(i)}_n e^{-\left(\frac{\mu}{q\delta(n+i+1)}-\frac{1}{2}(n+i+1)\delta\right)x}
      \left(1-qe^{-\delta x}\right)^{n+i+1} \nonumber \\
  &\quad \times P_n^{\left(\frac{2\mu}{q\delta^2(n+i+1)}-n-i-1, -\frac{2\mu}{q\delta^2(n+i+1)}-n-i-1\right)}
       \left(\frac{1+qe^{-\delta x}}{1-qe^{-\delta x}}\right), \\
  N^{(i)}_n &= N_{n+i}, 
\end{align}
respectively.\par
%
%
\section{More general extended deformed Hulth\'en potentials}

Instead of using the ground-state wavefunction $\phi_0(r)$ of the Eckart potential $U_{A,B}(r;\alpha)$ Schr\"odinger equation with energy ${\cal E}_0$ as factorization function, we plan to employ some polynomial-type, nodeless solution $\eta(r)$ of the same with energy ${\cal E}$ less than ${\cal E}_0$, in terms of which the superpotential will be written as $W(r) = - d\log\eta(r)/dr$ and the two partner potentials will become
\begin{equation}
  U^{(0)}(r) = W^2 - \frac{dW}{dr} + {\cal E}, \qquad U^{(1)}(r) = W^2 + \frac{dW}{dr} + {\cal E}.
\end{equation}
In Ref.~\cite{cq}, it was shown that, apart from some exceptional cases, such solutions $\eta(r)$ belong to three different types $\eta^{\rm I}_{A,B,m}$, $\eta^{\rm II}_{A,B,m}$, and $\eta^{\rm III}_{A,B,m}$, which are written explicitly in appendix A.\par
%
%
It turns out that to get for $U^{(1)}(r)$ some rationally-extended Eckart potential with given parameters $A$ and $B$, we have to start from some Eckart potential $U^{(0)}(r)$ with some different $A'$, but the same $B$. The results read \cite{cq}
\begin{align}
  U^{(0)}(r) &= U_{A',B}(r;\alpha), \nonumber \\
  U^{(1)}(r) &= U_{A,B,{\rm ext}}(r;\alpha) = U_{A,B}(r;\alpha) + U_{A,B,{\rm rat}}(r;\alpha), \nonumber \\
  U_{A,B,{\rm rat}}(r;\alpha) &= 2\alpha^2 (1-z^2) \Biggl\{2z \frac{\dot{g}^{(A,B)}_m}{g^{(A,B)}_m}
      - (1-z^2) \Biggl[\frac{\ddot{g}^{(A,B)}_m}{g^{(A,B)}_m} - \Biggl(\frac{\dot{g}^{(A,B)}_m}{g^{(A,B)}_m}
      \Biggr)^2\Biggr] - m\Biggr\},
\end{align}
where a dot denotes a derivative with respect to $z=\coth\alpha r$. According to the choice made for the factorization function $\eta(r)$, we may distinguish the three cases
\begin{eqnarray}
  & ({\rm I}) \; &  A' = A-\alpha, \quad \eta = \eta^{\rm I}_{A-\alpha,B,m}, \quad g^{(A,B)}_m(z) = 
         P^{(\alpha_m, \beta_m)}_m(z),  \nonumber \\
  && \alpha_m = \frac{1}{\alpha}\left[-A-(m-1)\alpha + \frac{B}{A+(m-1)\alpha}\right], \nonumber \\ 
  && \beta_m = \frac{1}{\alpha}\left[-A-(m-1)\alpha - \frac{B}{A+(m-1)\alpha}\right],  \nonumber \\
  && m = 1, 2, 3, \ldots, \quad A \ge 2\alpha, \quad (A-\alpha)^2 < B < (A-\alpha)[A+(m-1)\alpha]; \\
  & ({\rm II}) \; &  A' = A+\alpha, \quad \eta = \eta^{\rm II}_{A+\alpha,B,m}, \quad g^{(A,B)}_m(z) = 
         P^{(-\alpha_{-m-1}, -\beta_{-m-1})}_m(z),  \nonumber \\
  && \alpha_{-m-1} = \frac{1}{\alpha} \left[-A+m\alpha + \frac{B}{A-m\alpha}\right], \nonumber \\
  && \beta_{-m-1} = \frac{1}{\alpha}\left[-A+m\alpha - \frac{B}{A-m\alpha}\right],  \nonumber \\
  && m = 1, 2, 3, \ldots, \quad \frac{1}{2}(m-1)\alpha < A < m\alpha, \quad B > (A+\alpha)^2; \\
  & ({\rm III}) \; &  A' = A+\alpha, \quad \eta = \eta^{\rm III}_{A+\alpha,B,m}, \quad g^{(A,B)}_m(z) = 
         P^{(-\alpha_{-m-1}, -\beta_{-m-1})}_m(z),  \nonumber \\
  && \alpha_{-m-1} = \frac{1}{\alpha}\left[-A+m\alpha + \frac{B}{A-m\alpha}\right], \nonumber \\
  && \beta_{-m-1} = \frac{1}{\alpha}\left[-A+m\alpha - \frac{B}{A-m\alpha}\right],  \nonumber \\
  && m = 2, 4, 6, \ldots, \quad A>m\alpha, \quad B > (A+\alpha)^2. 
\end{eqnarray}
The first two are isospectral because $\eta^{-1}(r)$ is not normalizable, whereas, in the third case, the partner has an extra bound state below the spectrum of ${\cal H}_0$, corresponding to $\eta^{-1}(r)$ that is now normalizable.\par
%
%
In type I case, the bound-state spectra are given by
\begin{equation}
  {\cal E}^{(0)}_n = {\cal E}^{(1)}_n = - [A+(n-1)\alpha]^2 - \frac{B^2}{[A+(n-1)\alpha]^2}, \qquad
  0 \le n < \frac{\sqrt{B}-A+\alpha}{\alpha},
\end{equation}
and the corresponding wavefunctions can be written as
\begin{equation}
\begin{split}
  \phi^{(0)}_n(r) = {\cal N}^{(0)}_n (z-1)^{\alpha_n/2} (z+1)^{\beta_n/2} P_n^{(\alpha_n,\beta_n)}(z), \\
  \phi^{(1)}_n(r) = {\cal N}^{(1)}_n \frac{(z-1)^{\alpha_n/2} (z+1)^{\beta_n/2}}{g^{(A,B)}_m(z)}
       y^{(A,B)}_{m+n-1}(z),  \label{eq:phi-0-1}
\end{split}
\end{equation}
where
\begin{equation}
\begin{split}
  \alpha_n &= \frac{1}{\alpha}\left[-A-(n-1)\alpha+\frac{B}{A+(n-1)\alpha}\right], \\
  \beta_n &= \frac{1}{\alpha}\left[-A-(n-1)\alpha-\frac{B}{A+(n-1)\alpha}\right],
\end{split}. \label{eq:alpha-beta-bis}
\end{equation}
and
\begin{align}
  y^{(A,B)}_{m+n-1}(z) &= - \frac{(A-\alpha)^2 [A+(n-1)\alpha]^2 - B^2}{(A-\alpha) [A+(n-1)\alpha)]^2}
       g^{(A,B)}_m(z) P^{(\alpha_n,\beta_n)}_{n-1}(z) \nonumber \\
  &\quad + \frac{(A-\alpha)^2 [A+(m-1)\alpha]^2 - B^2}{(A-\alpha) [A+(m-1)\alpha]^2} g^{(A+\alpha,B}
       _{m-1}(z) P^{(\alpha_n,\beta_n)}_n(z). 
\end{align}
Note that, in (\ref{eq:phi-0-1}), ${\cal N}^{(0)}_n$ is given by eq.~(\ref{eq:Eckart-norm}), with $\alpha_n$ and $\beta_n$ defined in (\ref{eq:alpha-beta-bis}), while ${\cal N}^{(1)}_n = {\cal N}^{(0)}_n/\left[{\cal E}^{(0)}_n - {\cal E}^{\rm I}_{A-\alpha,B,m}\right]^{-1/2}$.\par
%
%
In type II case, the bound-state spectra read
\begin{equation}
  {\cal E}^{(0)}_n = {\cal E}^{(1)}_n = - [A+(n+1)\alpha]^2 - \frac{B^2}{[A+(n+1)\alpha]^2}, \qquad
  0 \le n <\frac{\sqrt{B}-A-\alpha}{\alpha},
\end{equation}
with corresponding wavefunctions similar to (\ref{eq:phi-0-1}), but with
\begin{equation}
\begin{split}
  \alpha_n &= \frac{1}{\alpha}\left[-A-(n+1)\alpha+\frac{B}{A+(n+1)\alpha}\right], \\
  \beta_n &= \frac{1}{\alpha}\left[-A-(n+1)\alpha-\frac{B}{A+(n+1)\alpha}\right],
\end{split}. \label{eq:alpha-beta-II}
\end{equation}
and $y^{(A,B)}_{m+n-1}(z)$ replaced by
\begin{align}
  y^{(A,B)}_{m+n+1}(z) &= - \frac{(A+\alpha)^2 [A+(n+1)\alpha]^2 - B^2}{(A+\alpha) [A+(n+1)\alpha)]^2}
       g^{(A,B)}_m(z) P^{(\alpha_n,\beta_n)}_{n-1}(z) \nonumber \\
  &\quad + \frac{(m+1)[2A-(m-1)\alpha]}{A+\alpha} g^{(A+\alpha,B}_{m+1}(z) P^{(\alpha_n,\beta_n)}_n(z). 
\end{align}
Here ${\cal N}^{(0)}_n$ is given by eq.~(\ref{eq:Eckart-norm}) with $\alpha_n$ and $\beta_n$ defined in (\ref{eq:alpha-beta-II}), while ${\cal N}^{(1)}_n = {\cal N}^{(0)}_n/\left[{\cal E}^{(0)}_n - {\cal E}^{(II)}_{A+\alpha, B,m}\right]^2$.\par
%
%
In type III case, the results are similar to those for type II, apart from the range of parameters and the existence of an extra bound state for ${\cal H}_1$ below ${\cal E}^{(1)}_0$. The latter is therefore the ground state with energy
\begin{equation}
  {\cal E}^{(1)}_{-m-1} = - (A-m\alpha)^2 - \frac{B^2}{(A-m\alpha)^2}
\end{equation}
and corresponding wavefunction
\begin{equation}
  \phi^{(1)}_{-m-1}(z) = {\cal N}^{(1)}_{-m-1} \frac{(z-1)^{\alpha_{-m-1}/2} (z+1)^{\beta_{-m-1}/2}}
  {g^{(A,B)}_m(z)},
\end{equation}
whose normalization coefficient has to be calculated separately.\par
%
%
On assuming now $A=(i+1)/2$, $B=\bar{\mu}/4$, and $\alpha=1/2$ in the previous results, we can obtain further extensions of the (already extended) Hulth\'en potentials $\bar{V}^{(i)}(r)$, defined in eqs.~(\ref{eq:def-V-bar-i}) and (\ref{eq:V-bar-i}),
\begin{equation}
  \bar{V}^{(i)}_{\rm ext}(r) = \bar{V}^{(i)}(r) + \bar{V}^{(i)}_{\rm rat}(r),
\end{equation}
where
\begin{align}
  \bar{V}^{(i)}_{\rm rat}(r) &= U_{\frac{i+1}{2}, \frac{\bar{\mu}}{4},{\rm rat}}\left(r;\frac{1}{2}\right)
       \nonumber \\
  &= \frac{1}{2}\left(1-\coth^2\,\frac{r}{2}\right) \Biggl\{2\coth\frac{r}{2} \frac{\dot{g}^{\left(\frac{i+1}{2},
       \frac{\bar{\mu}}{4}\right)}_m}{g^{\left(\frac{i+1}{2}, \frac{\bar{\mu}}{4}\right)}_m} \nonumber \\
  &\quad - \left(1-\coth^2\,\frac{r}{2}\right)\Biggl[\frac{\ddot{g}^{\left(\frac{i+1}{2},
       \frac{\bar{\mu}}{4}\right)}_m}{g^{\left(\frac{i+1}{2}, \frac{\bar{\mu}}{4}\right)}_m} -
       \Biggl(\frac{\dot{g}^{\left(\frac{i+1}{2},
       \frac{\bar{\mu}}{4}\right)}_m}{g^{\left(\frac{i+1}{2}, \frac{\bar{\mu}}{4}\right)}_m}\Biggr)^2
       \Biggr] - m \Biggr\}.
\end{align}
\par
%
%
In type I case, we get
\begin{equation}
  g^{\left(\frac{i+1}{2},\frac{\bar{\mu}}{4}\right)}_m(z) = P_m^{\left(-i-m+\frac{\bar{\mu}}{i+m},
  -i-m-\frac{\bar{\mu}}{i+m}\right)}\left(\coth\frac{r}{2}\right), 
\end{equation}
where
\begin{equation}
  m=1,2,3,\ldots, \qquad i\ge 1, \qquad i^2 < \bar{\mu} < i(i+m),
\end{equation}
and the results for bound states read
\begin{equation}
  \bar{E}^{(i,{\rm ext})}_n = \bar{E}^{(i-1)}_n = - \frac{1}{4} \left(\frac{\bar{\mu}-(n+i)^2}{n+i}\right)^2,
        \qquad 0 \le n <\sqrt{\bar{\mu}}-i, 
\end{equation}
\begin{align}
  \bar{\psi}^{(i,{\rm ext})}_n(r) &= \bar{N}^{(i,{\rm ext})}_n \frac{\left(\coth\frac{r}{2}-1\right)^{\frac{1}{2}
        \left(-i-n+\frac{\bar{\mu}}{i+n}\right)} \left(\coth\frac{r}{2}+1\right)^{\frac{1}{2}
        \left(-i-n-\frac{\bar{\mu}}{i+n}\right)}}{g^{\left(\frac{i+1}{2},\frac{\bar{\mu}}{4}\right)}_m\left(
        \coth\frac{r}{2}\right)} \nonumber \\ 
   &\quad \times y^{\left(\frac{i+1}{2},\frac{\bar{\mu}}{4}\right)}_{m+n-1}\left(\coth\frac{r}{2}\right),
\end{align}
\begin{align}
  & y^{\left(\frac{i+1}{2},\frac{\bar{\mu}}{4}\right)}_{m+n-1}\left(\coth\frac{r}{2}\right) \nonumber \\ 
  & \quad = - \frac{i^2(i+n)^2 - \bar{\mu}^2}{2i(i+n)^2} g^{\left(\frac{i+1}{2},\frac{\bar{\mu}}{4}\right)}_m
        \left(\coth\frac{r}{2}\right) P_{n-1}^{\left(-i-n+\frac{\bar{\mu}}{i+n},
        -i-n-\frac{\bar{\mu}}{i+n}\right)}\left(\coth\frac{r}{2}\right) \nonumber \\
  & \quad\quad + \frac{i^2(i+m)^2 - \bar{\mu}^2}{2i(i+m)^2} 
        g^{\left(\frac{i+2}{2},\frac{\bar{\mu}}{4}\right)}_{m-1}\left(\coth\frac{r}{2}\right) 
        P_n^{\left(-i-n+\frac{\bar{\mu}}{i+n},-i-n-\frac{\bar{\mu}}{i+n}\right)}\left(\coth\frac{r}{2}\right).           
\end{align}
\par
%
%
In type II and III cases, we obtain
\begin{equation}
  g^{\left(\frac{i+1}{2},\frac{\bar{\mu}}{4}\right)}_m(z) = P_m^{\left(i+1-m-\frac{\bar{\mu}}{i+1-m},
  i+1-m+\frac{\bar{\mu}}{i+1-m}\right)}\left(\coth\frac{r}{2}\right), 
\end{equation}
where
\begin{equation}
\begin{split}
  m &= 2,3,4,\ldots, \qquad \frac{1}{2}(m-3) < i < m-1, \qquad \bar{\mu} > (i+2)^2 \qquad \text{for type II}, \\
  m &= 2,4,6,\ldots, \qquad i > m-1, \qquad \bar{\mu} > (i+2)^2 \qquad \text{for type III}. 
\end{split}
\end{equation}
In both cases, the results for bound states with nonnegative $n$ read
\begin{equation}
  \bar{E}^{(i,{\rm ext})}_n = \bar{E}^{(i+1)}_n = - \frac{1}{4} \left(\frac{\bar{\mu}-(n+i+2)^2}
        {n+i+2}\right)^2,
        \qquad 0 \le n <\sqrt{\bar{\mu}}-i-2, 
\end{equation}
\begin{align}
  \bar{\psi}^{(i,{\rm ext})}_n(r) &= \bar{N}^{(i,{\rm ext})}_n \frac{\left(\coth\frac{r}{2}-1\right)^{\frac{1}{2}
        \left(-i-n-2+\frac{\bar{\mu}}{i+n+2}\right)} \left(\coth\frac{r}{2}+1\right)^{\frac{1}{2}
        \left(-i-n-2-\frac{\bar{\mu}}{i+n+2}\right)}}{g^{\left(\frac{i+1}{2},\frac{\bar{\mu}}{4}\right)}_m\left(
        \coth\frac{r}{2}\right)} \nonumber \\ 
   &\quad \times y^{\left(\frac{i+1}{2},\frac{\bar{\mu}}{4}\right)}_{m+n+1}\left(\coth\frac{r}{2}\right),
\end{align}
\begin{align}
  & y^{\left(\frac{i+1}{2},\frac{\bar{\mu}}{4}\right)}_{m+n+1}\left(\coth\frac{r}{2}\right) \nonumber \\ 
  & = - \frac{(i+2)^2(i+n+2)^2 - \bar{\mu}^2}{2(i+2)(i+n+2)^2} 
        g^{\left(\frac{i+1}{2},\frac{\bar{\mu}}{4}\right)}_m\left(\coth\frac{r}{2}\right) 
        P_{n-1}^{\left(-i-n-2+\frac{\bar{\mu}}{i+n+2},
        -i-n-2-\frac{\bar{\mu}}{i+n+2}\right)}\left(\coth\frac{r}{2}\right) \nonumber \\
  & \quad + \frac{(m+1)(2i-m+3)}{i+2} 
        g^{\left(\frac{i+2}{2},\frac{\bar{\mu}}{4}\right)}_{m+1}\left(\coth\frac{r}{2}\right) 
        P_n^{\left(-i-n-2+\frac{\bar{\mu}}{i+n+2},-i-n-2-\frac{\bar{\mu}}{i+n+2}\right)}\left(\coth\frac{r}
        {2}\right).           
\end{align}
In type III case, the extra bound state corresponds to
\begin{equation}
  \bar{E}^{(i,{\rm ext})}_{-m-1} = - \frac{1}{4} \left(\frac{\bar{\mu}-(i-m+1)^2}{i-m+1}\right)^2,
\end{equation}
\begin{align}
  \bar{\psi}^{(i,{\rm ext})}_{-m-1}(r) &= \bar{N}^{(i,{\rm ext})}_{-m-1} 
        \frac{\left(\coth\frac{r}{2}-1\right)^{\frac{1}{2}
        \left(-i-1+m+\frac{\bar{\mu}}{i+1-m}\right)} \left(\coth\frac{r}{2}+1\right)^{\frac{1}{2}
        \left(-i-1+m-\frac{\bar{\mu}}{i+1-m}\right)}}{g^{\left(\frac{i+1}{2},\frac{\bar{\mu}}{4}\right)}_m\left(
        \coth\frac{r}{2}\right)}.
\end{align}
It is worth observing that an extension of a pure deformed Hulth\'en potential (corresponding to $i=0$) can be obtained in type II case for $m=2$.
\par
%
%
It now remains to perform the changes of variable (\ref{eq:r-x}) and of parameter (\ref{eq:def-V-bar}) to obtain a generalization of Schr\"odinger equation (\ref{eq:gen-SE}),
\begin{equation}
  \left[- \frac{1}{2} \frac{d^2}{dx^2} + V^{(i,{\rm ext})}_q(x) - E^{(i,{\rm ext})}_n\right]
  \psi^{(i,{\rm ext})}_n(x) = 0,
\end{equation}
where
\begin{equation}
  V^{(i,{\rm ext})}_q(x) = V^{(i)}_q(x) + V^{(i,{\rm rat})}_q(x),
\end{equation}
with $V^{(i)}_q(x)$ given in eq.~(\ref{eq:gen-hulthen}).\par
%
%
Instead of providing the general expressions for $V^{(i,{\rm rat})}_q(x)$, which are rather complicated, we will list some examples together with the corresponding spectrum:
\begin{eqnarray}
  &\text{(i)} &\text{Type I with $m=1$, $i\ge 1$, and $\tfrac{1}{2}q\delta^2 i^2 < \mu < \tfrac{1}{2}
      q\delta^2 i(i+1)$} \nonumber \\
  & & V^{(i,{\rm rat})}_q(x) = - [i^2 (i+1)^2 q^2 \delta^4 - 4\mu^2] q\delta^2 e^{-\delta x} \nonumber \\
  & & \quad \times\{i(i+1)q\delta^2 - 2\mu + [i(i+1)q\delta^2 + 2\mu]q e^{-\delta x}\}^{-2}, \\
  & & E^{(i,{\rm ext})}_n = -\frac{1}{2} \left(\frac{\mu}{q\delta(n+i)} - \frac{\delta}{2}(n+i)\right)^2,
      \quad 0 \le n < \sqrt{\frac{2\mu}{q\delta^2}} - i; \\
  &\text{(ii)} &\text{Type I with $m=2$, $i=1$, and $\tfrac{1}{2}q\delta^2 < \mu < \tfrac{3}{2}q\delta^2$} 
      \nonumber \\
  & & V^{(1,{\rm rat})}_q(x) = -2\delta^2 (9q^2\delta^4 - 4\mu^2) \nonumber \\
  & & \quad \times [(3q\delta^2-2\mu) (6q\delta^2-2\mu) qe^{-\delta x} + 2(6q\delta^2-2\mu)
      (6q\delta^2+2\mu) q^2 e^{-2\delta x} \nonumber \\
  & & \quad + (3q\delta^2+2\mu) (6q\delta^2+2\mu) q^3 e^{-3\delta x}] \nonumber \\
  & & \quad \times [(3q\delta^2-2\mu) (6q\delta^2-2\mu) + 2(3q\delta^2-2\mu) (3q\delta^2+2\mu)
      qe^{-\delta x} \nonumber \\
  & & \quad + (3q\delta^2+2\mu) (6q\delta^2+2\mu) q^2 e^{-2\delta x}]^{-2}, \\
  & & E^{(1,{\rm ext})}_n = -\frac{1}{2} \left(\frac{\mu}{q\delta(n+1)} - \frac{\delta}{2}(n+1)\right)^2,
      \quad 0 \le n < \sqrt{\frac{2\mu}{q\delta^2}} - 1; \\
  &\text{(iii)} & \text{Type II with $m=2$, $i=0$, and $\mu > 2q\delta^2$} \nonumber \\
  & & V^{(0,{\rm rat})}_q(x) = - \frac{2\mu}{q^2\delta^2} (q^2\delta^4 - 4\mu^2) \nonumber \\
  & & \quad \times [2(q\delta^2+2\mu) (q^2\delta^4 - 2\mu q\delta^2 + 4\mu^2) qe^{-\delta x}
      + 2\mu (q^2\delta^4 - 20\mu^2) q^2 e^{-2\delta x} \nonumber \\
  & & \quad - 2(q\delta^2-2\mu) (q^2\delta^4 + 4\mu q\delta^2 + 8\mu^2) q^3 e^{-3\delta x}
      + 2\mu (q\delta^2-2\mu) (q\delta^2+2\mu) q^4 e^{-4\delta x}] \nonumber \\
  & & \quad \times [2\mu (q\delta^2+2\mu) + 2(q\delta^2-2\mu) (q\delta^2+2\mu) qe^{-\delta x}
      \nonumber \\ 
  & & \quad - 2\mu (q\delta^2-2\mu) q^2 e^{-2\delta x}]^{-2}, \\
  & & E^{(0,{\rm ext})}_n = -\frac{1}{2} \left(\frac{\mu}{q\delta(n+2)} - \frac{\delta}{2}(n+2)\right)^2,
      \quad 0 \le n < \sqrt{\frac{2\mu}{q\delta^2}} - 2; \\
  &\text{(iv)} &\text{Type III with $m=2$, $i=2$, and $\mu > 8q\delta^2$} \nonumber \\
  & & V^{(2,{\rm rat})}_q(x) = - 2\delta^2 (225q^2\delta^4 - 4\mu^2) \nonumber \\
  & & \quad\times [(12q\delta^2-2\mu) (15q\delta^2-2\mu) qe^{-\delta x} + 2(12q\delta^2-2\mu)
      (12q\delta^2+2\mu) q^2 e^{-2\delta x} \nonumber \\
  & & \quad + (12q\delta^2+2\mu) (15q\delta^2+2\mu) q^3 e^{-3\delta x}] \nonumber \\
  & & \quad \times [(12q\delta^2-2\mu) (15q\delta^2-2\mu) + 2(15q\delta^2-2\mu) (15q\delta^2+2\mu)
      qe^{-\delta x} \nonumber \\
  & & \quad + (12q\delta^2+2\mu) (15q\delta^2+2\mu) q^2 e^{-2\delta x}]^{-2}, \\
  & & E^{(2,{\rm ext})}_n = -\frac{1}{2} \left(\frac{\mu}{q\delta(n+4)} - \frac{\delta}{2}(n+4)\right)^2,
      \nonumber \\
  & & \quad \text{$n=-3$ and $0 \le n < \sqrt{\frac{2\mu}{q\delta^2}} - 4$}.  
\end{eqnarray}
\par%
%
\section{Conclusion}

In the present paper, by using a known relation between the deformed Hulth\'en potential $V_q(x)$ and the Eckart one, we have shown that the bound-state wavefunctions of the former can be written in terms of Jacobi polynomials and their normalization coefficient can be expressed in terms of a double finite sum.\par
%
%
The shape invariance property of the Eckart potential in standard SUSYQM, combined with such a relation, has then allowed us to easily rederive the set of extended deformed Hulth\'en potentials $V^{(i)}_q(x)$, $i=1$, 2,~\ldots, previously obtained by using the Darboux-Crum transformation \cite{hall18}, and to show that the spectra and normalized wavefunctions of the latter follow without any further calculations.\par
%
%
We would like to stress that the present approach considerably simplifies the calculation of the normalized eigenfunctions of the deformed Hulth\'en potential and of its simplest extensions, as provided in Ref.~\cite{hall18}.\par
%
%
{}Furthermore, we have constructed novel extensions $V^{(i,{\rm ext})}_q(x) = V^{(i)}_q(x) + V^{(i,{\rm rat})}_q(x)$ of the deformed Hulth\'en potential by taking advantage of some results previously obtained for rationally-extended Eckart potentials in a more general SUSY framework \cite{cq}. Such extended deformed Hulth\'en potentials belong to three different types, the first two isospectral to $V^{(i-1)}_q(x)$ or $V^{(i+1)}_q(x)$, respectively, and the third one with an extra bound state below the spectrum of $V^{(i+1)}_q(x)$.\par
%
%
As a final point, it is worth observing that, in contrast with the first set of extensions $V^{(i)}_q(x)$, whose partner in standard SUSYQM is simply $V^{(i+1)}_q(x)$, in the two isospectral cases, the second set $V^{(i,{\rm ext})}_q(x)$ enjoys, as the corresponding extended Eckart potentials, what has been called an enlarged shape invariance property in Ref.~\cite{cq}: together with the substitution of $i+1$ for $i$, the degree $m$ of the polynomial arising in the definition of $V^{(i,{\rm rat})}_q(x)$ has to be changed into $m-1$ or $m+1$ for type I or II, respectively.\par
%
%
\bigskip
\noindent
This work was supported by the Fonds de la Recherche Scientifique - FNRS under Grant Number 4.45.10.08.\par
%
%
\section*{\boldmath Appendix: Explicit expressions for $\eta(r)$}

\renewcommand{\theequation}{A.\arabic{equation}}
\setcounter{equation}{0}

In Ref.~\cite{cq}, the Eckart potential and its extensions have been dealt with under the assumption that $\alpha=1$. To get results for any positive value of $\alpha$, applicable to the deformed Hulth\'en potential, can be achieved by performing changes of variable and of parameters. On starting from
\begin{equation}
  \left[- \frac{d^2}{d\bar{r}^2} + \bar{U}_{\bar{A},\bar{B}}(\bar{r};1) - \bar{{\cal E}}_n\right] \bar{\phi}_n(\bar{r})
  = 0,
\end{equation}
where
\begin{equation}
  \bar{U}_{\bar{A},\bar{B}}(\bar{r};1) = - \bar{A}(\bar{A}-1) \operatorname{csch}^2 \bar{r} - 2\bar{B}
  \coth \bar{r},
\end{equation}
\begin{equation}
  \bar{{\cal E}}_n = - (\bar{A}+n)^2 - \frac{\bar{B}^2}{(\bar{A}+n)^2},
\end{equation}
eqs.~(\ref{eq:Eckart-pot}), (\ref{eq:Eckart-SE}), and (\ref{eq:Eckart-E}) can indeed be obtained by assuming
\begin{align}
  \bar{r} &= \alpha r, \qquad \bar{A} = \frac{A}{\alpha}, \qquad \bar{B} = \frac{B}{\alpha^2}, \qquad
       \bar{U}_{\bar{A},\bar{B}}(\bar{r};1) = \frac{1}{\alpha^2} U_{A,B}(r;\alpha), \nonumber \\
  \bar{{\cal E}}_n &= \frac{{\cal E}_n}{\alpha^2}, \qquad \bar{\phi}_n(\bar{r}) \propto \phi_n(r).
\end{align}
\par
%
%
On using such transformations, the polynomial-type, nodeless solutions of eq.~(\ref{eq:Eckart-SE}) with ${\cal E} < {\cal E}_0 = - A^2 - B^2/A^2$, can be written as
\begin{align}
  \eta^{\rm I}_{A,B,m}(r) &= \chi^{\rm I}_{A,B,m}(z) P_m^{\left(\frac{1}{\alpha}\left(-A-m\alpha
      +\frac{B}{A+m\alpha}\right), \frac{1}{\alpha}\left(-A-m\alpha-\frac{B}{A+m\alpha}\right)\right)}(z),
      \nonumber \\
  &\quad \text{if $m=1,2,3,\ldots$, $A\ge \alpha$, $A^2 < B < A(A+m\alpha)$}, \\
  \eta^{\rm II}_{A,B,m}(r) &= \chi^{\rm II}_{A,B,m}(z) P_m^{\left(\frac{1}{\alpha}\left[A-(m+1)\alpha
      - \frac{B}{A-(m+1)\alpha}\right], \frac{1}{\alpha}\left[A-(m+1)\alpha+ \frac{B}{A-(m+1)\alpha}\right]
      \right)}(z),  \nonumber \\
  &\quad \text{if $m=1,2,3,\ldots$, $\tfrac{1}{2}(m+1)\alpha < A < (m+1)\alpha$, $B > A^2$}, \\
   \eta^{\rm III}_{A,B,m}(r) &= \chi^{\rm III}_{A,B,m}(z) P_m^{\left(\frac{1}{\alpha}\left[A-(m+1)\alpha
      - \frac{B}{A-(m+1)\alpha}\right], \frac{1}{\alpha}\left[A-(m+1)\alpha+ \frac{B}{A-(m+1)\alpha}\right]
      \right)}(z),  \nonumber \\
  &\quad \text{if $m=2,4,6,\ldots$, $A > (m+1)\alpha$, $B > A^2$},
\end{align}
where $z = \coth \alpha r$,
\begin{align}
  \chi^{\rm I}_{A,B,m}(z) &= (z-1)^{-\frac{1}{2\alpha}\left(A+m\alpha-\frac{B}{A+m\alpha}\right)}
      (z+1)^{-\frac{1}{2\alpha}\left(A+m\alpha+\frac{B}{A+m\alpha}\right)}, \\
  \chi^{\rm II}_{A,B,m}(z) &= \chi^{\rm III}_{A,B,m}(z) \nonumber \\
  &= (z-1)^{\frac{1}{2\alpha}\left[A-(m+1)\alpha-\frac{B}{A-(m+1)\alpha}\right]}
      (z+1)^{\frac{1}{2\alpha}\left[A-(m+1)\alpha+\frac{B}{A-(m+1)\alpha}\right]},
\end{align}
and corresponding energies
\begin{align}
  {\cal E}^{\rm I}_{A,B,m} &= - (A+m\alpha)^2 - \frac{B^2}{(A+m\alpha)^2}, \\
  {\cal E}^{\rm II}_{A,B,m} &= {\cal E}^{\rm III}_{A,B,m} = - [A-(m+1)\alpha]^2 - \frac{B^2}
       {[A-(m+1)\alpha]^2}.
\end{align}
\par
%
%

\end{document}